\documentclass[journal]{IEEEtran} 
\IEEEoverridecommandlockouts

\usepackage{mathrsfs}
\usepackage[noadjust]{cite}
\usepackage{graphicx,color,overpic,psfrag}
\usepackage{amsmath, amssymb}
\usepackage{latexsym}
\usepackage{bm}
\usepackage{amssymb}
\usepackage{cases}
\usepackage{array}
\usepackage{fancyhdr}
\usepackage{setspace}
\usepackage{subcaption}
\usepackage{caption}

\UseRawInputEncoding
\usepackage{indentfirst}
\usepackage{cases}
\usepackage{url}
\usepackage{algpseudocode}
\usepackage{algorithm}
\usepackage{blkarray}

\usepackage{booktabs}
\usepackage{multirow}
\usepackage{dsfont}
\usepackage{tabularx}
\usepackage[table]{xcolor}
\usepackage{amsfonts}
\usepackage{amsthm}
\usepackage{stfloats}
\usepackage{letltxmacro}
\usepackage{lettrine}

\newcommand{\ra}[1]{\renewcommand{\arraystretch}{#1}}

\graphicspath{{figure/}}
\allowdisplaybreaks

\begin{document}

	\title{Channel Charting-assisted Non-orthogonal Pilot Allocation for Uplink XL-MIMO  Transmission}

	\author{ Haohong~Che, Li~You, Jue~Wang, Zhenzhou Jin, Chenjie Xie, 
		and Xiqi~Gao

		\thanks{

			Haohong Che, Li You, Zhenzhou Jin, Chenjie Xie, and Xiqi Gao are with the National Mobile Communications Research Laboratory, Southeast University, Nanjing 210096, China, and also with the Purple Mountain Laboratories, Nanjing 211100, China (e-mail: haohong\_che@seu.edu.cn; lyou@seu.edu.cn; zzjin@seu.edu.cn; cjxie@seu.edu.cn;  xqgao@seu.edu.cn).
		
			Jue Wang is with the School of Information Science and Technology, Nantong University, Nantong 226019, China, and also with the Nantong Research Institute for Advanced Communication Technologies, Nantong 226019, China
			(e-mail: wangjue@ntu.edu.cn).
		}
	}

	\maketitle
	
	\begin{abstract}
	Extremely large-scale multiple-input multiple-output (XL-MIMO) is critical to future wireless networks. The substantial increase in the number of base station (BS) antennas introduces near-field propagation effects in the wireless channels, complicating channel parameter estimation and increasing pilot overhead. Channel charting (CC) has emerged as a potent unsupervised technique to effectively harness varying high-dimensional channel statistics to enable non-orthogonal pilot assignment and reduce pilot overhead. In this paper, we investigate near-field channel estimation with reduced pilot overhead by developing a CC-assisted pilot scheduling. To this end, we introduce a polar-domain codebook to capture the power distribution of near-field XL-MIMO channels. The CC-assisted approach uses such features as inputs to enable an effective low-dimensional mapping of the inherent correlation patterns in near-field user terminal (UT) channels. Building upon the mapped channel correlations, we further propose a near-field CC-assisted pilot allocation (NCC-PA) algorithm, which efficiently enhances channel orthogonality among pilot-reusing UTs. Numerical results confirm that the NCC-PA algorithm substantially elevates the wireless transmission performance, offering a marked improvement over the conventional far-field CC-PA approach.
		
	\end{abstract}
	
	
\section{Introduction} \label{sec:introduction}
Advancing from existing massive multiple-input multiple-output (MIMO) technology, extremely large-scale MIMO (XL-MIMO) integrates extremely large aperture arrays to deliver significant improvement in spectral efficiency and energy efficiency \cite{NFdai2023}. As such, XL-MIMO emerges as a cornerstone technology for sixth-generation (6G) wireless networks, which is crucial for satisfying the stringent requirements of ubiquitous and ultra-dense connectivity \cite{LDMA}.

However, increasing the number of antennas introduces new channel characteristics, e.g., near-field propagation effects \cite{Rainbow, 2024XJ}. Within the near-field region, the wavefront propagates as spherical waves, diverging from the planar waves commonly assumed in the far-field \cite{2023CEdai, channelmea}. This substantially escalates the number of channel parameters requiring estimation, leading to high pilot overhead \cite{NFdai2023}. Limited pilot resources have become a bottleneck in channel acquisition, spurring research into reducing system pilot overhead \cite{2023CEdai, 2016PR}. Non-orthogonal pilot allocation, leveraging user terminals' (UTs') azimuth angles, has been recognized as an effective strategy to enhance performance \cite{UTG, LA-PA}. However, such non-orthogonal pilot allocation schemes exploit only the angular-domain differences of UT channels, neglecting the near-field effects and channel orthogonality arising from the sparsity of UT channels in the angle-distance domain.

The expanded aperture of XL-MIMO antenna arrays also enhances the spatial resolution of UT channels, allowing the statistical channel state information (s-CSI) (and CSI) to capture more detailed features strongly correlated with the propagation (or scattering) environment \cite{2023CC, jingdm4mmimo}. Significant correlation in the steady-state characteristics observed between two UT channels may imply proximity in their physical locations \cite{2021CS}. To better utilize such s-CSI (or CSI), channel charting (CC) has emerged as a promising framework that enables BSs to discern radio geometry and facilitate radio resource management \cite{2021CS}. Utilizing environmentally dependent steady-state channel characteristics, CC serves as a tool for unsupervised mapping of UT neighborhoods, ensuring that UTs with similar low-dimensional virtual coordinates have similar channel characteristics. Such mapping allows the BS to reconstruct a concise radio environment map for scheduling tasks, including pilot allocation. However, existing research predominantly relies on the far-field assumption, failing to leverage the near-field characteristics of XL-MIMO channels for CC-assisted pilot scheduling \cite{2023CC, Jin2024}.

In this paper, we investigate near-field channel estimation with reduced pilot overhead and formulate the non-orthogonal pilot allocation problem by exploiting the channel spatial correlations. We introduce the CC technique and detail the process of extracting spatial features from near-field UT channels, establishing a near-field CC (NCC) method. Then, we design an NCC-assisted algorithm to efficiently derive non-orthogonal pilot allocation schemes without the knowledge of UT locations. Via simulations, it is validated that the proposed algorithm enhances channel estimation performance compared to the far-field CC (FCC) method.



\section{System Model and Problem Statement}

We focus on the uplink (UL) channel training phase in XL-MIMO transmission, where a BS with \( M \) antennas serves \( K \left(K\ll \textit{M}\right)\) single-antenna UTs. The set \( \mathcal{K} = \{1, 2, \ldots, K\} \) enumerates the indices of UTs. To describe the near-field propagation effects in XL-MIMO wireless channels, the UL channel from UT $k$ to the BS is modeled as \cite{2023CEdai}:
\begin{equation}\label{eq:1}
	\mathbf{h}_k=\iint_{\mathfrak{D}}\bm{b}\left(\theta,\rho\right)h_k\left(\theta,\rho\right)\text{d} \theta\text{d}\rho
\end{equation}	
where $\mathfrak{D}$ denotes the range of the near-field region of the BS antenna array, and $h_k(\theta, \rho)$ represents the complex-valued channel gain function of UT $k$, while the BS array response vector $\bm{b}\left(\theta,\rho\right)\in \mathbb{C}^{M \times 1}$ is obtained based on the spherical-wave assumption \cite{NFdai2023}.   

With the channel model in (\ref{eq:1}),  spatial correlation in near-field channels depends not only on the angles at which signals arrive, but also on the locations of the UTs (or scatters), which is further described by the power location spectrum (PLS) \cite{2022ZY}. We assume that the channels at different locations are independent of each other, i.e., $\mathbb{E}\{h\left(\theta,\rho\right)h^*\left(\theta',\rho'\right)\}=\xi f\left(\theta,\rho\right)\delta\left(\theta-\theta'\right)\delta\left(\rho-\rho'\right)$, where $f\left(\theta,\rho\right)$ and $\xi$ denote the channel PLS and the large-scale fading coefficient, respectively. Hence, the near-field channel covariance matrix of UT $k$, denoted as $\bm{\Phi}_k$, can be described as \cite{2016PR, 2022ZY}:
\begin{equation}
	\begin{aligned}
		\bm{\Phi}_k&=\mathbb{E}\{\mathbf{h}_k\mathbf{h}_k^H\} \\
		&= \xi _k\iint_{\mathfrak{D}} \bm{b}\left(\theta,\rho\right)\bm{b}\left(\theta,\rho\right)^H f\left(\theta,\rho \right) \text{d}\theta \text{d}\rho.
	\end{aligned}
\end{equation}

To reduce the system pilot overhead, we propose non-orthogonal pilot allocation to admit intra-cell pilot reuse \cite{LA-PA, youliPR}. Briefly, we define a set $\mathcal{T}=\{1, 2, \ldots, \tau\}$ of $\tau$ orthogonal pilot sequences, where $\tau$ denotes the pilot length. We employ the sum mean square error (MSE), denoted as $\varepsilon=\sum_{\substack{k=1}}^{K}\varepsilon_k$, to quantify the channel estimation performance. The channel estimation error for UT $k$ can be expressed as \cite{youliPR}:
\begin{equation}\label{eq3}
	\varepsilon_k=\text{tr}\left\{\bm{\Phi}_k-\bm{\Phi}_k \left(\sum_{l,k\in \mathcal{P}_t, l\neq k}\bm{\Phi}_l+\frac{1}{\zeta\tau}\mathbf{I}\right)^{-1}\bm{\Phi}_k\right\}.
\end{equation}
where set $\mathcal{P}_t\left(t\in \mathcal{T}\right)$ and $\zeta$ denote the group of UTs reusing the pilot sequence $t$ and the signal-to-noise ratio (SNR) for UL channel training, respectively. Since channel covariance matrices can be accurately estimated with sufficient time-frequency resources \cite{youliPR, 2016PR}, we assume that $\bm{\Phi}_k\left(\forall k\in \mathcal{K}\right)$ can be obtained at the BS.

Based on (\ref{eq3}), it is straightforward to deduce that: for any two UTs $k, k'\in \mathcal{K}$, the minimum value of $\varepsilon$ is attainable only when $\Vert\bm{\Phi}_k\bm{\Phi}_{k'}^H\Vert_2=0$ if they utilize the same pilot \cite{2016PR}. Thus, we formulate the non-orthogonal pilot allocation problem for UL channel training in XL-MIMO systems by exploiting the spatial correlations between UT channels as follows:
\begin{subequations} \label{eq6} 
	\begin{align}
		\text{P1}: \min_{\mathcal{P}_t} \quad &\sum_{t=1}^{\tau} \sum_{\substack{k,k' \in \mathcal{P}_t, k < k'}} \left\| \bm{\Phi}_k \bm{\Phi}_{k'}^H \right\|_2  \label{P1main}\\
		\text{s.t.} \quad &\bigcup_{t=1}^{\tau} \mathcal{P}_t = \mathcal{K}, \label{P1.1} \\
		& \mathcal{P}_t \neq \varnothing, \quad \forall t \in \mathcal{T}, \label{P1.2}   \\
		& \mathcal{P}_t \cap \mathcal{P}_{t'} = \varnothing, \quad \text{for } t' \in \mathcal{T}, t \leq t', \label{P1.3} 
	\end{align}
\end{subequations}
where the constraints (\ref{P1.1}), (\ref{P1.2}), and (\ref{P1.3}) are imposed to guarantee that each UT employs one pilot sequence and all pilot sequences are assigned.

\section{NCC-assisted Non-orthogonal Pilot Allocation}
In this section, we address problem P1 by implementing an NCC-assisted pilot allocation strategy. Using the virtual coordinate distances between UTs on the CC map to represent channel correlations, we transform the objective of minimizing channel correlation among pilot-reusing UTs in \text{P1} into maximizing the distance between the virtual coordinates of such UTs. Then, we propose an NCC-assisted pilot allocation (NCC-PA) algorithm engineered to derive pilot allocation schemes efficiently.

\subsection{Channel Charting}
To better exploit scattering environment-related information in UT channels, we extract the spatial features from the UT channel covariance matrices collected at the BS, rather than relying on instantaneous CSI. Utilizing such extracted spatial features, we develop an NCC method to reconstruct the UTs' local neighborhoods and channel correlation patterns for further pilot scheduling.

\subsubsection{Overview of CC}
CC depends on extracting spatial features of UTs from high-dimensional s-CSI (or CSI) and determines the relative positions of points on a map using a specific forward charting function \(\mathbb{f}\), such as manifold learning \cite{2023CC, 2021CS}, ensuring that UTs with similar coordinates on the map also have similar spatial features, which is defined as follows:
\begin{equation}
	\mathbb{f}: \bm{C} \leftarrow \Xi
\end{equation}
where $\Xi=\left[\bm{e}_1,\cdots,\bm{e}_K\right]$ and $\bm{C}=\left[\bm{c}_1,\cdots,\bm{c}_K\right] \in \mathbb{R}^{\mathcal{D} \times K}$ denote the extracted features of UT channel and chart coordinates of each point on the map, respectively, and $\mathcal{D}$ represents the dimension of CC map.

\begin{figure*}[t]
	\centering
	\begin{tabular}{ccc} 
		\includegraphics[scale=0.35]{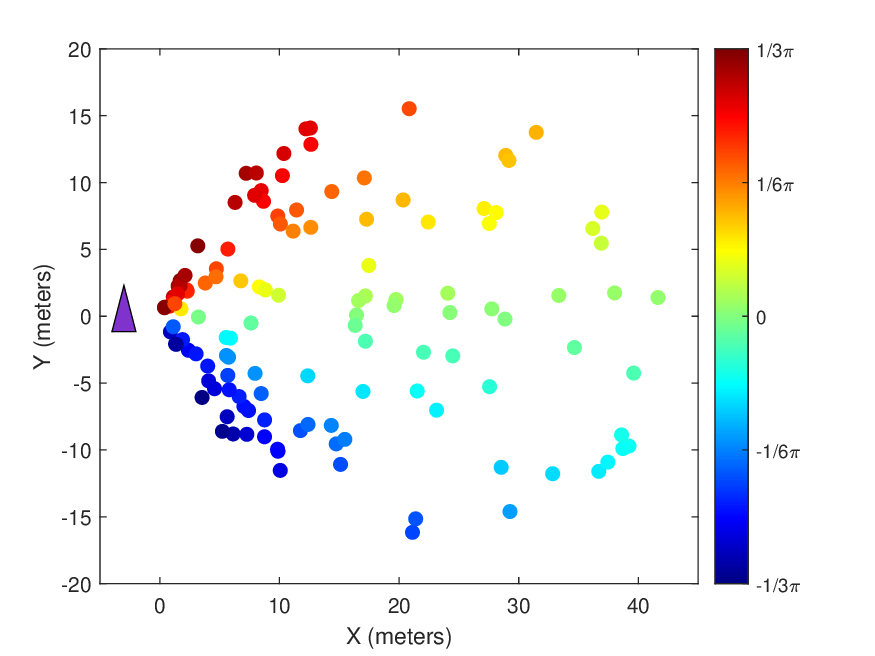} & 
		\includegraphics[scale=0.35]{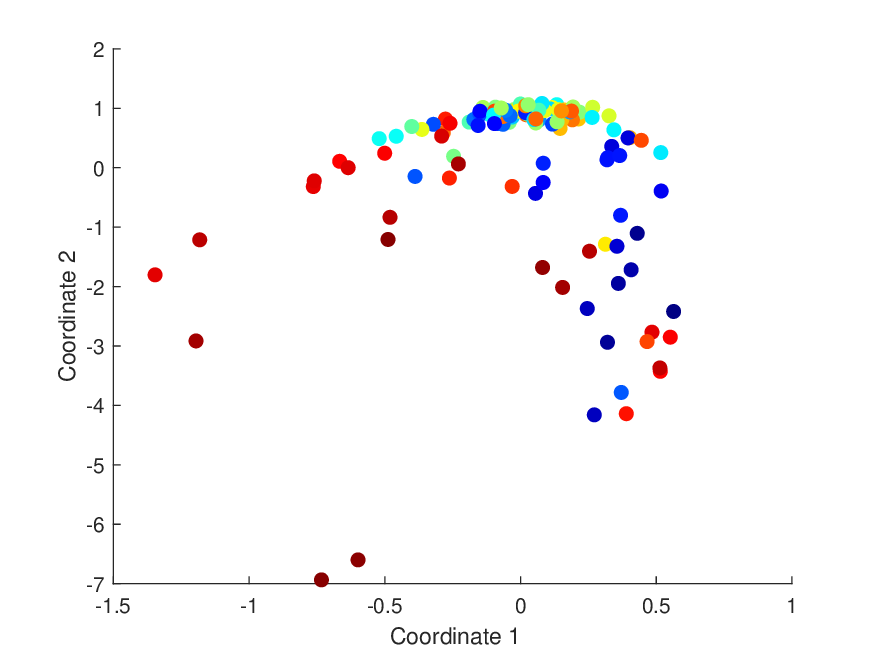} &
		\includegraphics[scale=0.35]{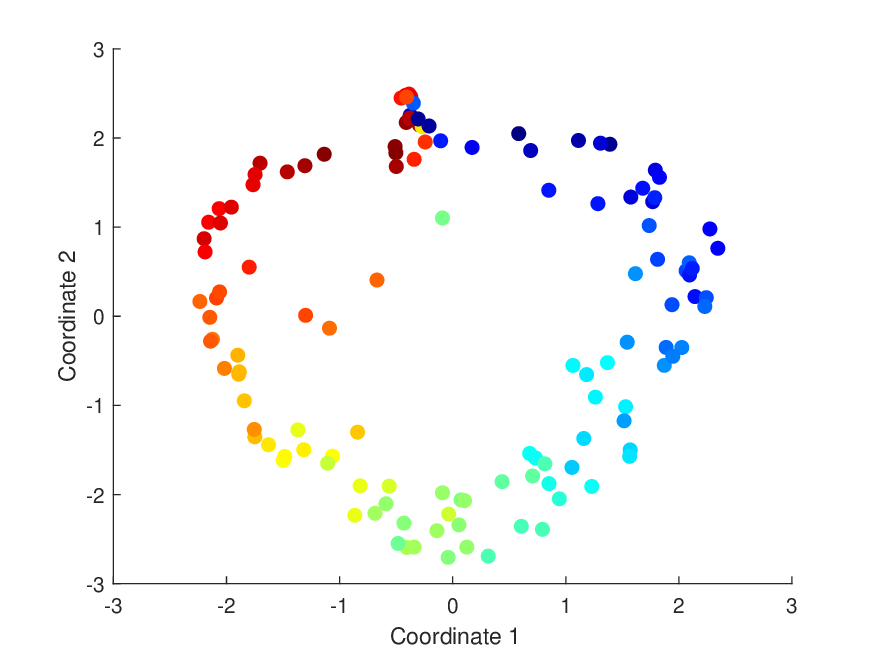} \\
		{\footnotesize\sf (a)} & {\footnotesize\sf (b)} & {\footnotesize\sf (c)} \\
	\end{tabular}
	\caption{Illustration of CC in an XL-MIMO setup with 128 UTs: (a) The purple triangle denotes the BS equipped with a 256-element ULA, where each colored marker indicates a UT. (b) A 2D CC map created using far-field features $\bm{F}$. (c) The same CC map using near-field features $\bm{\Xi}$.}
	\label{fig:CC}
\end{figure*}

\subsubsection{Feature Extraction}
An initial feature extraction step is designed to transform raw s-CSI into a format more amenable to analysis \cite{2023CC}. The extracted steady-state channel characteristics, closely linked to the propagation environment, can be seen as functions of UT-specific scattering information on a manifold \cite{2023CC, 2021CS}. Thus, the correlation of such characteristics likely reflects UTs' physical proximity, making the constructed CC more accurately represent real neighboring relationship.

Given that the BS cannot obtain precise instantaneous CSI of UTs prior to pilot scheduling, this part focuses on extracting spatial features from the UT channel covariance matrices obtained by the BS to serve as input for CC, enabling the reconstruction of UTs' local neighborhoods \cite{2021CS, 2022CCPA}. For this purpose, we present a near-field polar domain codebook $\bm{W}$ to facilitate a more detailed decomposition of UT channel covariance matrices, which is described as \cite{2023CEdai}:
\begin{equation}
	\bm{W}=\left[\bm{W}_1,\bm{W}_2,\bm{W}_3,\cdots,\bm{W}_{S}\right]  \in \mathbb{C}^{M\times MS}
\end{equation}
where $\bm{W}_s=\left[\bm{b}\left(\theta_1,\rho^{\theta_1}_{s}\right),\cdots,\bm{b}\left(\theta_M,\rho^{\theta_M}_{s}\right) \right]$ represents a matrix comprised of $M$ steering vectors, with $\theta_m=\frac{2m-M+1}{M}  \left(m=0,1,\cdots,M-1\right)$ representing angular uniform sampling and $\rho^{\theta}_{s}=\frac{M^2d^2}{2s\lambda_c\beta_{\triangle}}\left(1-\theta^2\right) \left(s=1,2,3,\cdots S\right)$ describing non-uniform sampling in terms of distance, $d$ and $\lambda_c$ denote antenna spacing and the wavelength of carrier $f_c$, respectively, $\beta_{\triangle}$ is a positive-valued moderator variable. 

Considering that the near-field channel manifests polar domain sparsity, and hence the steering vectors with various array configurations exhibit asymptotic orthogonality in the near field \cite{NFdai2023, LDMA}, the UT channel covariance matrix can subsequently be represented as \cite{2016PR}:
\begin{equation}\label{eq:approx}
	\begin{aligned}
		\bm{\Phi}_k&=\mathbb{E}\{\mathbf{h}_k\mathbf{h}_k^H\} \\
		&=
		\sum_{s=1}^{S} \xi_k \int_{-\pi}^{\pi} \bm{b}\left(\theta,\rho^{\theta}_{s}\right)\bm{b}\left(\theta,\rho^{\theta}_{s}\right)^Hf_k\left(\theta,\rho^{\theta}_{s}\right)\text{d}\theta \\
		&=\sum_{s=1}^{S}\bm{W}_s\bm{r}_s^k\bm{W}_s^H  = \bm{W}\bm{\Lambda}_{NF}^k\bm{W}^H 
	\end{aligned}
\end{equation}
where $\bm{\Lambda}_{NF}^k=\text{diag}\{\text{vec}\{\bm{\Gamma}_k\}\}$ and $\bm{\Gamma}_k=\left[\bm{r}_1^k,\cdots,\bm{r}_S^k\right]\in \mathbb{C}^{M\times S}$ contain the polar-domain PLS information of the UT channels. The elements of $\bm{r}_{s}^k\in \mathbb{C}^{M\times 1}$ can be represented as (\ref{eq:lambda}), shown at the bottom of this page.

\begin{figure*}[b]
	\hrule
	\vspace{0.2cm} 
	\begin{equation}
		\label{eq:lambda}
		\left[\bm{r}^k_s\right]_m=\xi _k M \cdot f_k\left(\theta_{m-1},\rho^{\theta_{m-1}}_{s-1} \right)\left(\theta_m-\theta_{m-1}\right)\left(\rho^{\theta_{m-1}}_{s}-\rho^{\theta_{m-1}}_{s-1}\right), \text{for} \quad m=1,2,\cdots,M.
	\end{equation}
\end{figure*}

Equation (\ref{eq:approx}) illustrates that within the near-field, the channel covariance matrices of UTs can be decomposed into a consistent structure.\footnote{The channel covariance matrices of far-field UTs can likewise be represented in a similar structure. When all UTs are within the far-field region, angular features alone are sufficient for spatial differentiation between UTs \cite{youliPR, 2022CCPA}. Under such conditions, $\bm{W}$ simplifies to an $M$-dimensional unitary discrete Fourier transform matrix $\bm{D}$, and $\bm{\Phi}_k$ for the UT $k$ can be expressed as $\bm{\Phi}_k=\bm{D}\text{diag}\{\bm{r}_k\}\bm{D}^H$ with $\bm{r}_k=\sum_{s=1}^{S} \bm{r}_s^k$.} For any UT \(k \in \mathcal{K}\), the diagonal matrix \(\bm{\Lambda}_{NF}^k\) is influenced by both the PLS \(f_k(\theta, \rho)\) and the large-scale fading effects associated with signal propagation, capturing the polar-domain distributions of near-field UTs.

Given that NCC aims to map the polar-domain neighboring relationship and correlation patterns between UT channels, cosine similarity, a metric for correlation assessment \cite{2021CS, youliPR}, is employed as the dissimilarity metric to extract spatial features from the covariance matrices of UT channels. This process can be expressed as follows \cite{2021CS}:
\begin{equation}\label{metric}
	\mathfrak{d}\left(\bm{e}_k, \bm{e}_{k'}\right) = 2 - \frac{2|\bm{e}_k^H \bm{e}_{k'}|}{\|\bm{e}_k\|_2 \|\bm{e}_{k'}\|_2},\quad \forall k,k'\in \mathcal{K}.
\end{equation}

Unlike the feature matrix $\bm{F}$ in conventional FCC method, which only attempts to extract angular domain information \cite{2022CCPA}, i.e., $\left[\bm{F}\right]_{ij}=\mathfrak{d}\left(\bm{r}_i, \bm{r}_j\right)$, in this part, polar-domain spatial features are extracted from the near-field channel covariance matrix of each UT and aggregated into a feature matrix $\bm{\Xi}$, with elements defined as $\left[\bm{\Xi}\right]_{ij}=\mathfrak{d}\left(\text{vec}\{\bm{\Gamma}_i\}, \text{vec}\{\bm{\Gamma}_j\}\right)$.

\subsubsection{Chart Coordinates}

Leveraging the obtained polar-domain spatial features \(\bm{\Xi}\), we propose an NCC approach to generate the CC map, as detailed in \textbf{Algorithm \ref{alg:1}}. In the NCC algorithm, Isometric mapping (Isomap), a dimensionality reduction (DR) technique grounded in manifold learning \cite{isomap}, is selected as the forward charting function \(\mathbb{f}\). This technique effectively preserves the geodesic distances among UT nodes, thereby ensuring that the Euclidean distances, defined as \(\bm{d}(\mathbf{a}, \mathbf{b}) = \sqrt{\sum_{i=1}^{n} (a_i - b_i)^2}\), on the CC maps accurately reflect the actual distance between UTs on the embedded low-dimensional manifold \cite{isomap}.

\begin{algorithm}\small
	
	\caption{Near-field Channel Charting (NCC)} \label{alg:1}
	\textbf{Input:} {UT channel covariance matrices $\bm{\Phi}_k\left(\forall k\in\mathcal{K}\right)$, CC dimension $\mathcal{D}$ }
	\begin{algorithmic}[1]
		\State \textbf{Feature extraction:} build feature matrix $\Xi \in \mathbb{R}^{K\times K }$ by $\left[\Xi\right]_{ij}=\mathfrak{d}\left(\bm{\Lambda}_{NF}^i,\bm{\Lambda}_{NF}^j\right)$
		
		\State \textbf{DR:}  $\bm{C}=\left[\bm{c}_1,\cdots,\bm{c}_K\right] \leftarrow \mathbb{f}\left(\Xi\right),\forall k\in \mathcal{K}$	
	\end{algorithmic}
	\textbf{Output:} {CC coordinates $\bm{C} \in \mathbb{R}^{\mathcal{D} \times K }$}
\end{algorithm}

To provide an intuitive explanation for our proposed NCC method, we present the CC maps for an example scenario in Fig. 1. Fig. 1(a) illustrates an environment where a 256-element uniform linear array (ULA) XL-MIMO BS serves 128 UTs, based on the near-field channel model specified in (\ref{eq:1}). The purple triangle represents the BS location, and the UTs are denoted as colored circles, with each color denoting its azimuth angles relative to the BS. Figs. 1(b) and 1(c) display the CC maps generated via Isomap, using feature matrices \(\bm{F}\) and \(\bm{\Xi}\), respectively, which correspond to the FCC and NCC feature extraction methods.

Based on the orthogonality assessment metric described in (\ref{metric}) \cite{youliPR}, the CC maps shown in Fig. 1 are designed to visualize channel correlations using different features, mapping UTs with stronger correlations in closer proximity. However, unlike Fig. 1(b),  which fails to distinguish UTs at different locations, Fig. 1(c) effectively maps neighboring UTs with similar azimuth angles to adjacent positions on the CC map, thus demonstrating the efficacy of the NCC method.

\subsection{PA Algorithm Design}\label{sec_alogrithm}

In this part, we apply the NCC approach to handle problem \text{P1}. Given that spatial correlations between channels are deduced from distances on the CC map, we base our strategy for allocating non-orthogonal pilots on the proximity of UTs. Note that UTs located closer on the CC map tend to experience stronger channel correlations and increased mutual interference. Therefore, minimizing the channel correlation among pilot-reusing UTs in \text{P1} is equivalent to maximizing the spread between their virtual coordinates. We have developed the NCC-PA algorithm to enhance channel estimation performance, which utilizes geodesic distances between UTs. As outlined in \textbf{Algorithm \ref{alg:2}}, this approach ensures maximal spatial separation between pilot-reusing UTs on the CC map.

As outlined in Step 4 of \textbf{Algorithm \ref{alg:2}}, using the NCC approach, pilot 1 is assigned to UT 1, designated as the central node \(k^*\), and distinct pilot sequences are allocated to the nearest \(\tau-1\) UT nodes relative to \(k^*\). Step 8 focuses on selecting the closest node from the remaining UTs to act as the new central node \(k^*\), thereby maximizing the average distance between UT nodes sharing the same pilot sequence. After all pilot sequences are assigned to UT nodes on the map, the BS employs a minimum MSE (MMSE) estimator for channel estimation, with performance assessed according to (\ref{eq3}).

\begin{algorithm}\small
	
	\caption{NCC-assisted Pilot Allocation (NCC-PA)} \label{alg:2}
	\textbf{Input:} {UT set $\mathcal{K} = \{1, 2,\cdots, K\}$, the CC coordinates $\bm{c}_k\left(\forall k\in \mathcal{K}\right)$, available pilot set $\mathcal{T}$}
	\begin{algorithmic}[1]
		\State \textbf{Initialization:} $\mathcal{P}_{1}=\{1\}$, $\mathcal{P}_{t}=\emptyset$ $(t=2 \cdots,\tau )$, then $\mathcal{K}^{un}=\mathcal{K}\setminus \{1\}$, $k^*=1$
		
		\While{$\mathcal{K}^{un}\neq \emptyset$}
		\For{$t =2:\tau$}
		\State Select the UT $k$ by 
		$\underset { k\in\mathcal{K}^{un}}  { \operatorname {arg\,min} } \, \bm{d}\left(\bm{c}_k,\bm{c}_{k^*}\right)$
		\State Allocate pilot sequence $t$ to UT $k$ 
		\State Update $\mathcal{P}_{t} \leftarrow \mathcal{P}_{t} \bigcup \{k\}$, $\mathcal{K}^{un} \leftarrow \mathcal{K}^{un}\setminus \{k\}$
		\EndFor
		\State Select the UT $k$ by 
		$\underset { k\in\mathcal{K}^{un}} { \operatorname {arg\,min} } \, \bm{d}\left(\bm{c}_k,\bm{c}_{k^*}\right)$
		\State Update $k^*\leftarrow k$
		\State Allocate pilot sequence $1$ to UT $k^*$
		\State Update $\mathcal{P}_{1} \leftarrow \mathcal{P}_{1} \bigcup \{k^*\}$, $\mathcal{K}^{un} \leftarrow \mathcal{K}^{un}\setminus \{k\}$
		\EndWhile
		
	\end{algorithmic}
	\textbf{Output:} {Pilot allocation schemes $\mathcal{P}_{t}\left(\forall t\in \mathcal{T}\right)$}
\end{algorithm}

Finally, the computational complexity of our proposed NCC-PA algorithm is evaluated as follows. During the NCC-PA algorithm, a maximum of $K^2$ vector inner products are computed, directly resulting in a computational complexity of $\mathcal{O}\left(\mathcal{D}K^2\right)$. Furthermore, the computational complexity of the Isomap is bounded by $\mathcal{O}\left(\left(MS+\text{log}K\right)K^2+K^3\right)$ \cite{2021CS}. Consequently, the overall computational complexity of our proposed NCC-PA algorithm is given by $\mathcal{O}\left(\left(MS+\text{log}K+\mathcal{D}\right)K^2+K^3\right)$.

\section{Simulation Results}\label{simu}
In this section, the performance of the proposed NCC-PA algorithm is evaluated through detailed simulations. We consider a single-cell XL-MIMO scenario serving 16 UTs with the carrier frequency of 30 GHz, and Table \ref{tb:simulation parameters} summarizes the simulation parameters. The BS is equipped with a ULA consisting of 256 antennas positioned at intervals of half a wavelength. All UTs are randomly distributed within the near-field region,\footnote{ With the BS equipped with \( M = 256 \) antennas and a carrier wavelength of \( \lambda_c = 0.01 \) meters, the spacing between antennas is set to \( \varrho = \lambda_c/2 = 0.005 \) meters. Consequently, the Rayleigh distance is computed as \( \left(2M^2\varrho^2\right)/\lambda_c \approx \left(M^2\lambda_c\right)/2 = 327.68 \) meters \cite{NFdai2023}. UTs located within the Rayleigh distance fall within the near-field region.} where the angles and distances between the UTs (or scatterers) and the BS follow distributions $ \mathcal{U}\left( -\pi/3, \pi/3 \right) $ and $ \mathcal{U}\left( 5, 70 \right)$ (in meter), respectively. As existing channel power spectrum models are not suitable for near-field channels, we have utilized the simultaneous orthogonal matching pursuit (SOMP) algorithm for polar-domain channel estimation of near-field UTs \cite{2023CEdai}, i.e., $\mathbf{h}^p=\bm{W}^H\mathbf{h}$, for channel power distribution calculations. We set $\beta_{\triangle}$ to 1.8, and the number of sampled distances on each angle is $S=12$. The matrix $\bm{\Lambda}_{NF}^k$ in (\ref{eq:approx}) is approximated by averaging the outcomes of 500 Monte Carlo simulations $\mathbf{h}^p\left(\mathbf{h}^p\right)^H$, where each UT is allocated $L=6$ paths \cite{2023CEdai}.\footnote{The average angle of arrival and distance parameters for each path are randomly generated within the near-field range. According to near-field channel measurements \cite{channelmea}, the azimuthal dispersion for different UTs is uniformly set at $30^\circ$.} For a fair comparison, we consider the following baselines:

\textit{Random-PA}: Pilots are assigned randomly.

\textit{FCC-PA}: Isomap is employed to implement CC. The far-field angular features $\bm{F}$, extracted from the near-field UT channel, serve as inputs. The nearest-neighbor PA algorithm \cite{2022CCPA} is utilized to conduct pilot scheduling.

\newcolumntype{L}{>{\hspace*{-\tabcolsep}}l}
\newcolumntype{R}{c<{\hspace*{-\tabcolsep}}}
\definecolor{lightblue}{rgb}{0.93,0.95,1.0}
\begin{table}[htbp]
	\captionsetup{font=footnotesize}
	\caption{Simulation Parameters}\label{tb:simulation parameters}
	\centering
	\ra{1.3}
	\scriptsize
	\begin{tabular}{LR}
		\toprule
		Parameter &  Value\\
		\midrule
		
		\rowcolor{lightblue}
		Carrier frequency & $30$ GHz \\
		Number of BS antennas & $ 256 $ \\
		\rowcolor{lightblue}	
		Number of UTs & $ 16 $ \\
		The distribution of path gain & $ \mathcal{CN}\left(0, 1\right) $  \\
		\rowcolor{lightblue}
		Number of angular domain samples &  $256$ \\
		Number of distance domain samples&  $12$ \\
		\rowcolor{lightblue}
		Number of channel paths & $6$\\
		Number of channel paths detected in SOMP & $12$\\
		\bottomrule
	\end{tabular}
\end{table}

\begin{figure}[htbp]
	\centering
	\includegraphics[scale=0.40]{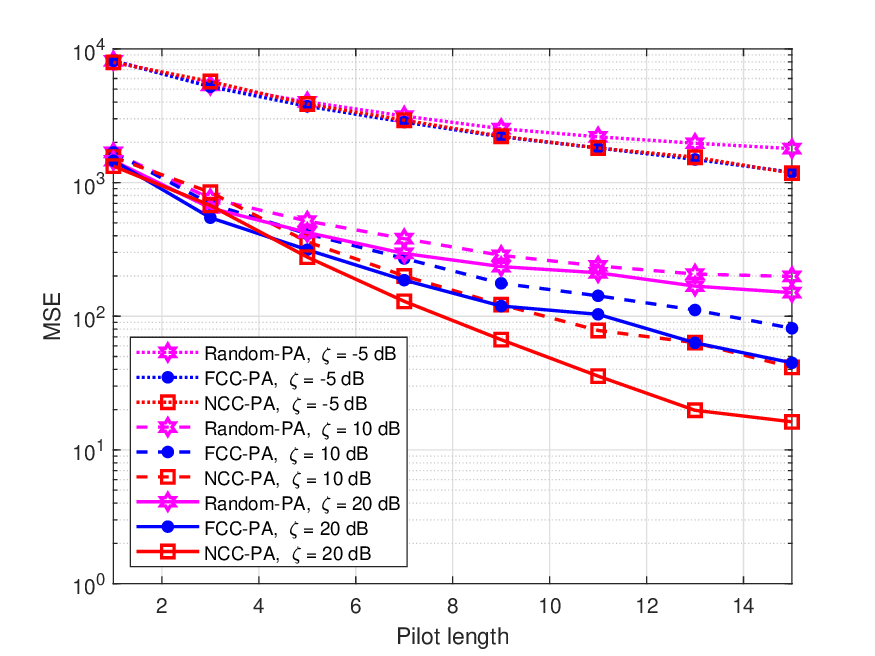}
	\captionsetup{font=footnotesize}
	\caption{MSE performance comparison against pilot length for different SNR values.}\label{fig:mmse}
\end{figure}

Firstly, the mean square error performance relative to pilot length is depicted in Fig. 2. As illustrated, the proposed NCC-PA algorithm, by harnessing the polar-domain spatial features of near-field UT channels $\bm{\Xi}$, delivers enhanced MSE performance compared to both the conventional Random-PA and FCC-PA algorithms, especially at longer pilot lengths. As SNR increases, the NCC-PA algorithm consistently demonstrates a marked superiority over the FCC-PA algorithm. This trend suggests that the NCC-PA effectively leverages the channel orthogonality arising from the sparsity in the angle-distance domain within the near-field range, thereby mitigating the interference among pilot-sharing UTs. Additionally, the disparity in MSE performance between the FCC-PA and NCC-PA algorithms widens with increasing SNR, further highlighting the effectiveness of the NCC-PA approach.

\begin{figure}[htbp]
	\centering
	\includegraphics[scale=0.40]{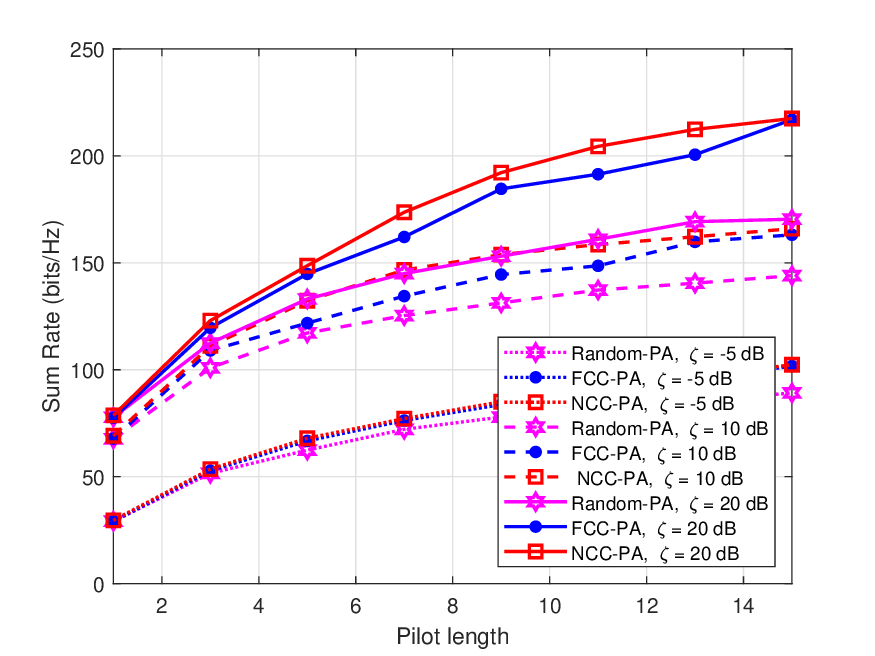}
	\captionsetup{font=footnotesize}
	\caption{Sum rate performance comparison against pilot length for different SNR values.}\label{fig:sumrate}
\end{figure}

The sum rate is derived from the data transmission framework outlined in \cite{youliPR}, which incorporates a robust MMSE receiver at the BS that compensates for channel estimation inaccuracies. Fig. 3 plots the sum rate performance versus the pilot length for different SNR levels. The results confirm that the proposed NCC-PA algorithm substantially exceeds the performance of the Random-PA and FCC-PA algorithms. The performance gains observed could be attributed to the proposed NCC-PA algorithm, which offers a more accurate representation of UT channel spatial correlations by leveraging spatial features in the polar domain, thereby significantly enhancing channel orthogonality among UTs assigned the same pilot sequence. Such enhancement reduces inter-user interference, leading to improved signal-to-noise and interference ratios, particularly in high-SNR scenarios. 




\section{Conclusion}\label{sec_conclusion} 

This paper investigated near-field pilot resource allocation strategies for XL-MIMO systems, emphasizing a CC-based approach. Initially, we introduce CC to tackle the task of non-orthogonal pilot scheduling in near-field communications. Subsequently, we extracted channel spatial features representing UT positional information from the covariance matrices of near-field UTs. We proposed an NCC approach to analyze the spatial correlation of near-field UT channels. Finally, we proposed an NCC-assisted PA algorithm to derive non-orthogonal pilot allocation schemes. Simulation results indicate that the proposed NCC approach substantially enhances the extraction of near-field spatial features from UT channels. By implementing the NCC approach, the MSE and sum rate performance of the proposed NCC-PA algorithm have substantially improved over the conventional FCC-based approach.

	\bibliographystyle{IEEEtran}
	\bibliography{EE_AI}
	
\end{document}